\documentclass[10pt]{article}

\usepackage[letterpaper]{geometry}

\usepackage{hicss}

\usepackage{accsupp}
\usepackage[export]{adjustbox}
\usepackage{amsmath}
\usepackage{array}
\usepackage[style=apa]{biblatex}
\usepackage{booktabs}
\usepackage{enumitem}
\usepackage{fontawesome5}
\usepackage{graphicx}
\usepackage[colorlinks=true, linkcolor=blue, citecolor=blue, urlcolor=blue]{hyperref}
\usepackage[none]{hyphenat}
\usepackage{indentfirst}
\usepackage{latexsym}
\usepackage{makecell}
\usepackage{minted}
\usepackage{multirow}
\usepackage{soul}
\usepackage{stfloats}
\usepackage{tikz}
\usepackage{times}
\usepackage{url}

\newcommand{\ic}[3]{\BeginAccSupp{ActualText={#3}}{\color{#2}#1}\EndAccSupp{}}

\newcommand{\rangeglyph}[3]{%
\begingroup
\pgfmathsetmacro{\pmed}{(#2-#1)/(#3-#1)}%
\begin{tikzpicture}[x=2.04cm,y=1ex,baseline=(base)]
  \draw[line width=0.45pt] (0,0) -- (1,0);

  \draw[line width=0.45pt] (0,-0.32) -- (0,0.32);
  \draw[line width=0.45pt] (1,-0.32) -- (1,0.32);

  \fill (\pmed,0) circle (1.15pt);

  \node[font=\footnotesize,anchor=south,inner sep=0pt,outer sep=0pt]
    at (\pmed,0.65) {#2};

  \node[font=\footnotesize,anchor=north west,inner sep=0pt,outer sep=0pt]
    at (0,-0.55) {#1};

  \node[font=\footnotesize,anchor=north east,inner sep=0pt,outer sep=0pt]
    at (1,-0.55) {#3};

  \coordinate (base) at (0,-1.35);
\end{tikzpicture}%
\endgroup
}

\newcommand{\rangeglyphcompact}[3]{%
\begingroup
\pgfmathsetmacro{\pmed}{(#2-#1)/(#3-#1)}%
\begin{tikzpicture}[x=0.75cm,y=1ex,baseline=-0.5ex]
  \draw[line width=0.9pt] (0,0) -- (1,0);
  \draw[line width=0.9pt] (0,-0.6) -- (0,0.6);
  \draw[line width=0.9pt] (1,-0.6) -- (1,0.6);
  \fill (\pmed,0) circle (1.9pt);
\end{tikzpicture}%
\endgroup
}

\graphicspath{{figures/}}

\usetikzlibrary{positioning,arrows.meta,calc,shapes.geometric,backgrounds,fit}

\title{The Token Tax of Epistemic Accuracy: Comparing RAG and Long-Context Architectures for Document-Grounded Generative AI Applications}

\author{
Austin Hamilton \\
Miami University \\ \\
Ibrahim Yousif \\
Miami University \\ \\
Mohammad Mayyas \\
Miami University
\And
Ryan Singh \\
Miami University \\ \\
Arthur Carvalho \\
Miami University \\ \\
Lora A. Cavuoto \\
University at Buffalo
\And
Michael Wise \\
Miami University \\ \\
Zhe Shan \\
Miami University \\ \\
Fadel M. Megahed \\
Miami University
\AND
\parbox{\textwidth}{\centering\small
Please direct correspondence to Arthur Carvalho (\ul{arthur.carvalho@miamioh.edu}) or Fadel Megahed (\ul{fmegahed@miamioh.edu}).
}
}

\date{June 17, 2026}

\begin{document}

\maketitle

\begin{abstract}
Document-grounded assistants built on large language models are increasingly used in high-stakes, knowledge-intensive work. 
Their usefulness, however, may depend on how evidence is allocated before generation. 
We investigate such a claim by comparing two grounding architectures: (a) retrieval-augmented generation (RAG) that retrieves a few relevant passages, and (b) long-context prompting, which loads the whole document collection in context. We view these as two regimes of ``epistemic access'' on an accuracy--cost frontier. 
We use ``epistemic accuracy'' to capture model correctness that depends on having the right evidence. We posit that broader access (via long context) can increase it, but with a ``token tax'' (i.e., a substantial increase in cost due to larger input token consumption). 
We probe this framing with a case study in manufacturing safety training. 
Using an expert-validated benchmark, we evaluate 972 answers across three machines, two small language models, and three retrieval/in-context prompting approaches. 
Long-context prompting achieved the highest correctness (73.1\% vs. 65.4\% for semantic RAG), but at 26 times the per-query token cost. 
We interpret this gap as the token tax of broader evidentiary access.
We carefully discuss the implications of our findings for resource-constrained organizations. 
\end{abstract}
\subsubsection*{Keywords:}

Artificial intelligence, epistemic accuracy, long-context prompting, retrieval-augmented generation, token economics.

\section{Introduction}
\label{sec:introduction}

Large language models (LLMs) power modern chatbots and agentic artificial intelligence (AI) applications that can answer questions about policies, procedures, manuals, code, and other institutional documents. 
In these settings, answer quality depends on two input parameters: (a) the evidence placed in the model context (``memory"), and (b) the model's ability to reason over this context and generate an appropriate answer to the question or prompt. 

Historically, the context windows of mainstream OpenAI models, such as GPT-3.5 and early GPT-4 variants, ranged from approximately 4,000 to 32,000 tokens, with GPT-3.5 commonly offering 4,000 or 16,000 contexts and GPT-4 initially available in 8,000 and 32,000 versions.
This made \textit{retrieval-augmented generation} (RAG) a natural approach, in which a small subset of text is retrieved and placed in the model context before generation \parencite{Lewis2020}. However, newer frontier models can support much larger context windows, enabling the inclusion of substantially larger document collections in context before generation. 
We refer to this latter approach as \textit{long-context prompting}.
For example, at the time of writing, OpenAI's most advanced model, GPT-5.5, supports a context window of up to 1.05 million tokens.

These two context management approaches establish distinct conditions for answer correctness. In RAG, the model can generate a correct answer only if the retrieval step provides the necessary evidence. Consequently, errors may arise if the retriever fails to identify the relevant passage, if the model misuses a retrieved passage, or if both stages are unsuccessful. Retrieval is particularly fragile for multi-hop questions that require synthesizing evidence from multiple sections, as a single retrieval step may retrieve only some of the necessary passages \parencite{folha2024towards,Gao2024}. Long-context prompting mitigates this explicit retrieval failure by exposing the model to a broader set of evidence. However, it does not eliminate the challenge of evidence selection; rather, it transfers this responsibility to the model's internal attention and reasoning mechanisms over a much larger context \parencite{Liu2024}. This approach also increases input-token consumption, resulting in a recurring cost for broader evidentiary access \parencite{bai2026ai}.

We use the term \emph{epistemic accuracy} to describe this evidence-dependent form of correctness, defined as the extent to which a document-grounded AI system produces a correct answer because the necessary evidence is both available and appropriately utilized. The qualifier \emph{epistemic} reflects the notion that certain errors can be reduced through improved information or evidence, analogous to the distinction between epistemic and aleatoric uncertainty \parencite{hullermeier2021aleatoric}. This perspective reframes the AI architecture decision as an accuracy–cost trade-off. 
In particular, broader evidentiary access may enhance epistemic accuracy by providing the model with more documentation. However, it also increases input-token consumption and the operational burden associated with larger-context systems \parencite{bai2026ai}. We refer to the additional input-token cost paid for broader evidentiary access, and therefore for the possibility of higher epistemic accuracy, as the \emph{token tax}.
This motivates our overarching research question:

\begin{quote}
\textit{To what extent does broader evidentiary access improve epistemic accuracy in document-grounded AI systems, and what costs does this improvement entail?}
\end{quote}

The constraints present in manufacturing safety training offer a practical example for examining how this trade-off manifests. 
Facilities may utilize different generations of machines such as manual mills, CNC machines, and collaborative robots. 
Each machine operates under standard operating procedures (SOPs), setup procedures, equipment manuals, and safety modes. Although essential information is codified, it is often dispersed across equipment manuals, SOPs, maintenance records, and safety protocols. AI assistants are beginning to address this gap by answering worker questions directly from such documentation \parencite{Singh2026}. 
However, the burden of designing and deploying such systems can be significant for small and medium-sized enterprises (SMEs), which constitute approximately 98\% of U.S. manufacturers \parencite{SBAOffice2025}. Manufacturing SMEs frequently operate with limited budgets, minimal safety staff, and a widening skills gap \parencite{DeloitteManufacturingInstitute2024}. 
This setting serves as a case study in our work to illustrate how broader evidentiary access can improve answer correctness while introducing a recurring token cost.

We argue that this is a relevant domain as the accuracy of generated answers is paramount in safety-critical environments. 
When an AI assistant provides responses based on SOPs and manuals, it influences how workers interpret stop conditions or hazards and, consequently, how they act. Posing such questions effectively delegates part of the epistemic task (namely, locating and interpreting relevant documentation) to an autonomous system \parencite{BairdMaruping2021,Agerfalk2020}. 
Beyond accuracy, \textcite{Singh2026} argued that the performance of manufacturing safety-training chatbots should also be evaluated based on \textit{latency} and \textit{cost}.

\begin{figure}[htbp]
\centering
\definecolor{ragfill}{HTML}{E8F1FF}
\definecolor{ragborder}{HTML}{2E6DA4}
\definecolor{lcfill}{HTML}{FDE7D6}
\definecolor{lcborder}{HTML}{C0392B}
\definecolor{hubfill}{HTML}{EFEFEF}
\definecolor{hubborder}{HTML}{444444}
\definecolor{consfill}{HTML}{E9F6EC}
\definecolor{consborder}{HTML}{2E8B57}
\definecolor{escfill}{HTML}{FFF3E0}
\begin{tikzpicture}[
  font=\footnotesize,
  >={Latex[length=1.8mm]},
  every node/.style={transform shape},
  box/.style={rectangle, draw, rounded corners=2.5pt, line width=0.5pt, align=center, inner sep=4pt, fill=white},
  wide/.style={box, text width=5.6cm, draw=hubborder},
  hub/.style={box, text width=5.6cm, fill=hubfill, draw=hubborder},
  rag/.style={box, text width=2.7cm, fill=ragfill, draw=ragborder},
  lc/.style={box, text width=2.7cm, fill=lcfill, draw=lcborder},
  cons/.style={box, text width=5.6cm, fill=consfill, draw=consborder},
  esc/.style={box, text width=5.6cm, fill=escfill, draw=hubborder},
  ar/.style={->, line width=0.6pt, draw=black!55},
]
\node[wide, text width=6cm] (worker) at (0,0) {\textbf{AI question}, e.g., a manufacturing safety question};
\node[hub] (hub) at (0,-1) {\textbf{AI (manufacturing safety) assistant}\\[1pt]\emph{retrieves} $+$ \emph{interprets} evidence};
\node[rag] (rag) at (-1.55,-2.3) {\textbf{RAG}\\[1pt]selective~\textbar~low cost};
\node[lc] (lc) at (1.55,-2.3) {\textbf{Long context}\\[1pt]exhaustive~\textbar~token tax};
\node[cons] (cons) at (0,-3.6) {\textbf{Grounded accuracy}~~vs.~~\textbf{token $+$ capability cost}};
\draw[ar] (worker) -- (hub);
\draw[ar] (hub.south) -- (rag.north);
\draw[ar] (hub.south) -- (lc.north);
\draw[ar] (rag.south) -- (rag.south |- cons.north);
\draw[ar] (lc.south) -- (lc.south |- cons.north);
\end{tikzpicture}

\vspace{-0.25\baselineskip}

\caption{Two regimes of epistemic delegation.}
\label{fig:delegation-regimes}
\end{figure}
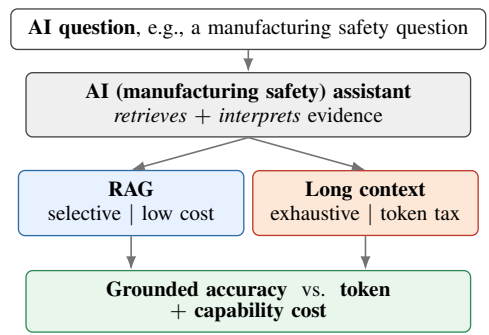

For SMEs, the choice of the context management architecture directly affects accuracy, latency, and cost. Specifically, this decision determines whether a firm opts for narrower and less expensive evidentiary access, which may overlook critical passages, or invests in broader access that may enhance correctness. The central consideration is not only whether an assistant can answer a safety question, but also whether the firm can sustain the chosen architecture. Figure~\ref{fig:delegation-regimes} summarizes this decision. RAG conserves context but increases the risk of retrieval failure, whereas long-context prompting broadens evidentiary access at the expense of higher token usage and cost.

Prior work has examined RAG \parencite{Lewis2020,Gao2024}, long-context models \parencite{Liu2024,bai2024longbench}, and SME digital readiness \parencite{hansen2021artificial,oldemeyer2025investigation} separately. Less is known about how grounding architecture mediates the cost of epistemic accuracy in resource-constrained, safety-relevant settings. 
Using an expert-validated benchmark of machine-specific safety questions and gold answers \parencite{Singh2026}, we compare semantic RAG, keyword RAG, and long-context prompting across two  language models of different sizes.
The benchmark covers three  manufacturing machines chosen to span legacy, automated, and human-robot equipment, including a Bridgeport manual mill, a Haas TL-1 CNC lathe, and a Universal Robots UR5e collaborative robot (cobot). 
Our results support our epistemic accuracy propositions by showing that long-context prompting achieved the highest correctness, albeit at a substantially higher cost.

Overall, our contributions are twofold. 
First, we suggest a conceptual framing of RAG and long-context prompting as two regimes of \emph{epistemic access}, with propositions describing their accuracy and cost in terms of token taxes.
Second, we illustrate the above framing through an empirical study in manufacturing safety training, showing that long-context prompting achieves higher correctness at a substantial token premium.

The remainder of the paper is organized as follows.
Section \ref{sec:background} reviews the relevant literature and discusses its applicability to manufacturing training environments. 
Section \ref{sec:methods} describes the research methodology, experimental design, benchmark, and evaluation metrics used to compare the two AI architectures. 
Section \ref{sec:results} presents the empirical results, analyzing the architectures across the dimensions of accuracy, cost, and latency. 
Finally, Section \ref{sec:discussion} discusses the implications of the findings, outlines limitations, and concludes with future research directions.

\section{Research background and related work}
\label{sec:background}

Grounding a model with retrieved text can enhance performance on knowledge-intensive tasks by providing relevant, up-to-date information \parencite{Lewis2020}.
This benefit is clear in many real-world settings, where grounded domain assistants have matched the performance of larger general-purpose models while being able to cite their sources \parencite{megahed2024introducing}. Still, a RAG answer depends on the quality of the passage the retriever finds. Sometimes, the retriever picks an irrelevant passage, returns text that is only loosely related but incorrect, or misses important information \parencite{Gao2024}. 
As such, changing the retrieval strategy alone can substantially impact accuracy \parencite{Singh2026}. 
The limitations of the RAG approach become most apparent when answering a question requires information to be combined from multiple passages. \parencite{folha2024towards}.
For example, a single retrieval step might find some relevant text but miss other key parts \parencite{Gao2024}.
In safety-critical situations, missing even one passage is more than a technical issue; it can change the instructions a worker follows. 
This makes retrieval the main barrier to epistemic accuracy, separate from the language model itself. 
A natural question that arises is when these retrieval problems can justify paying more for a different approach.

The more expensive option is long-context prompting.
This approach loads the entire document collection relevant to a task, so there is no risk of missing a passage because everything is included. 
However, research shows that this does not solve the bigger challenge, which is getting the model to use the right passage. 
Models do not read long inputs evenly, e.g., they pay less attention to information in the middle of their context \parencite{Liu2024}.
Some direct comparisons show that long-context prompting can sometimes outperform RAG when the model has enough resources. However, the extra context information might not always make a difference \parencite{li2024retrieval}. 
Retrieval that keeps the order of documents can even do better than long-context prompting while using fewer tokens \parencite{yu2024defense}. 

Cost is the other side of the trade-off, and it mainly depends on the number of input tokens.
Long-context prompting uses the most tokens by loading the whole data collection \parencite{bai2026ai}. 
This cost comes with every query, whether paid to a cloud provider or through on-premise hardware \parencite{pan2025cost}. In short, long context gives broader access by letting the model choose the evidence, but it means paying a higher token cost each time, which we refer to as the token tax.
Previous research has not directly compared this trade-off in a resource-limited setting, such as those experienced by SMEs.

What is being traded here is a specific kind of correctness. Some errors cannot be avoided, but others happen because of missing knowledge and could be fixed with better evidence \parencite{hullermeier2021aleatoric}. Grounding architectures target this second, fixable type of error, which is what we mean by epistemic accuracy, i.e., correctness depends on the evidence provided, and it improves when the right evidence is available \parencite{farquhar2024detecting}.
As stated before, previous research explains how to measure grounded correctness, but it has not compared how different architectures balance correctness and cost.

Whether this trade-off is worthwhile depends on who is making the decision. For small and medium-sized manufacturers, ongoing costs are the main concern. Studies on small-business systems show that successful adoption depends more on the owner's skills and limited resources than on the technology itself \parencite{thong1999integrated}. Because these companies operate with tight budgets, a failed investment is hard to recover from. Reviews of AI and Industry 4.0 adoption highlight the same issues: limited expertise, high costs, and weak infrastructure are the biggest barriers \parencite{oldemeyer2025investigation}. Many of these firms do not meet the basic requirements of standard maturity models \parencite{mittal2018critical}, and most choose ready-made tools because they lack the expertise to build custom solutions \parencite{hansen2021artificial}. Importantly, research separates one-time costs from ongoing maintenance, and risk-averse firms are most concerned about recurring costs, like per-query token fees \parencite{ghobakhloo2022drivers}. The resource-based view explains why these costs affect firms differently: it is the company’s own skills and resources, not just the tool, that determine whether adoption succeeds \parencite{caldeira2003using,MikalefGupta2021}.

\begin{figure}[h!]
\centering
\definecolor{ragfill}{HTML}{E8F1FF}
\definecolor{ragborder}{HTML}{2E6DA4}
\definecolor{lcfill}{HTML}{FDE7D6}
\definecolor{lcborder}{HTML}{C0392B}
\definecolor{qfill}{HTML}{EFEFEF}
\definecolor{ansfill}{HTML}{E9F6EC}
\begin{tikzpicture}[
  font=\footnotesize,
  pbox/.style={rectangle, draw=black!45, rounded corners=2pt, line width=0.5pt,
               align=center, inner sep=3pt, fill=white,
               text width=3.25cm, minimum height=0.5cm},
  ar/.style={->, line width=0.5pt, draw=black!55},
  lname/.style={align=center, font=\footnotesize, inner sep=1pt},
  note/.style={align=center, font=\footnotesize, text=black!70,
               inner sep=1pt, text width=4cm},
]
\def\xL{-2.05}
\def\xR{2.05}

\node[pbox, fill=qfill, text width=5cm] (q) at (0,0)
   {\ic{\faHardHat}{black!70}{Worker}~\textbf{Worker's question}\\{\scriptsize\color{black!60} e.g.\ a safety or setup step}};

\node[pbox] (r1) at (\xL,-2.05) {\ic{\faLayerGroup}{ragborder}{Document chunk index}~\textbf{Chunk index}\\{\scriptsize\color{black!60} manuals split into searchable pieces}};
\node[pbox] (r2) at (\xL,-3.25) {\ic{\faSearch}{ragborder}{Retrieve the top k passages}~\textbf{Retrieve top-$k$}\\{\scriptsize\color{black!60} pull the most relevant passages}};
\node[pbox] (r3) at (\xL,-4.35) {\ic{\faRobot}{ragborder}{Language model}~\textbf{LLM}\\{\scriptsize\color{black!60} reads only those passages}};
\node[pbox, fill=ansfill] (r4) at (\xL,-5.35) {\ic{\faQuoteRight}{ragborder}{Grounded answer with citation}~\textbf{Grounded answer}\\{\scriptsize\color{black!60} with citations to a document}};

\node[pbox] (l1) at (\xR,-2.05) {\ic{\faBook}{lcborder}{All manuals loaded in full as context}~\textbf{Full context}\\{\scriptsize\color{black!60} every manual loaded in full}};
\node[pbox] (l2) at (\xR,-3.40) {\ic{\faRobot}{lcborder}{Language model}~\textbf{LLM}\\{\scriptsize\color{black!60} reads all manuals at once}};
\node[pbox, fill=ansfill] (l3) at (\xR,-4.95) {\ic{\faCheckCircle}{lcborder}{Synthesised answer}~\textbf{Synthesised answer}\\{\scriptsize\color{black!60} drawn from across the manuals}};

\node[lname, anchor=north west, text=ragborder] (hrag) at (-3.65,-0.75)
   {\textbf{RAG}\\{\scriptsize retrieve, augment, then generate}};
\node[lname, anchor=north east, text=lcborder] (hlc) at (3.65,-0.75)
   {\textbf{Long context}\\{\scriptsize whole corpus in prompt}};

\node[draw=ragborder, dotted, line width=0.7pt, rounded corners=4pt,
      fit=(hrag)(r4), inner xsep=3pt, inner ysep=3pt] (ragbox){};
\node[draw=lcborder, dotted, line width=0.7pt, rounded corners=4pt,
      fit=(hlc)(l3), inner xsep=3pt, inner ysep=3pt] (lcbox){};

\node[note, anchor=north] at (ragbox.south)
   {Low token cost.\\
    \ic{\faExclamationTriangle}{orange!80!black}{Risk}~\emph{may miss a relevant passage}};
\node[note, anchor=north] at (lcbox.south)
   {High token cost: the \emph{token tax}.\\
    \ic{\faExclamationTriangle}{orange!80!black}{Risk}~\emph{evidence can be lost in a long context}};

\draw[ar] (q) -- (r1);
\draw[ar] (q) -- (l1);
\draw[ar] (r1) -- (r2);
\draw[ar] (r2) -- (r3);
\draw[ar] (r3) -- (r4);
\draw[ar] (l1) -- (l2);
\draw[ar] (l2) -- (l3);
\end{tikzpicture}

\vspace{-0.4\baselineskip}

\caption{Inside the two regimes of Figure~\ref{fig:delegation-regimes}.}
\label{fig:rag-architecture}
\end{figure}
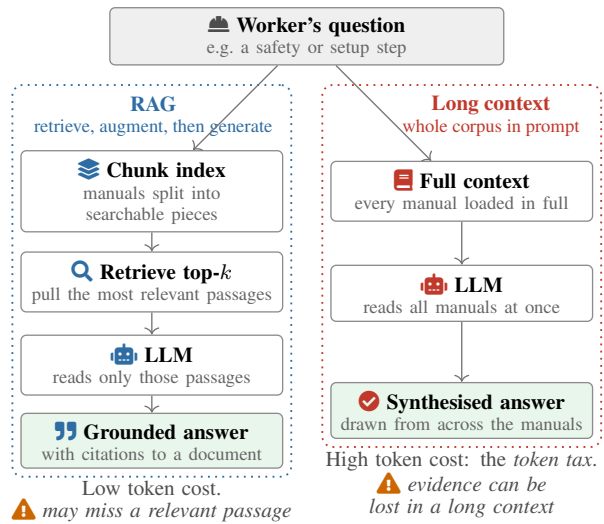

Manufacturing safety training puts both alternatives to the test. 
The information workers need is available, but it is spread out across manuals, procedures, and standards. A lot of this information is not in regular text; important details are found in tables, diagrams, icons, connector layouts, and warning labels \parencite{Singh2026}.
Sometimes, answering a question means piecing together evidence from different parts of a manual. 
For example, a worker might ask, ``What happens if my robot stopped after 1000 ms when my stopping-time safety limit is set to 200 ms?'' To answer this, one needs to connect the safety-limit definition, the stopping-time behavior, and how the control system interprets both. 
A wrong answer about a robot stop mode, emergency stop, or lockout procedure can affect how a worker handles a real hazard. 
To the best of our knowledge, a direct comparison of grounding architectures in these realistic conditions remains missing.
The above discussion leads to the conclusion that RAG and long-context prompting are different ways of choosing what evidence an LLM uses to generate an answer and how that evidence is selected.
Figure~\ref{fig:rag-architecture} summarizes the main differences between both approaches.

 \section{Methodology}
\label{sec:methods}

We next describe the three main components of our experimental methodology: the AI solutions evaluated, the benchmark used for comparison, and the metrics used to assess performance.

\subsection{AI solutions}

We compare three approaches, grouped into two architectures. 
The first architecture is RAG, represented by solutions built on OpenAI \textit{semantic retrieval} and OpenAI \textit{keyword retrieval}. 
Semantic retrieval searches for passages that are conceptually similar to the user query, even when the query and the searched documents use different words.
Keyword retrieval searches for passages that contain the same or closely matching terms as the query. 
We selected these two RAG approaches because they were the strongest-performing RAG configurations in the manufacturing-related experiments reported by \textcite{Singh2026}.
The second architecture is long-context prompting, represented by a prompt that includes all three manuals simultaneously. 
Table \ref{tab:architectures} summarizes the studied solutions.

\vspace{-.5\baselineskip}

\begin{table}[htb!]
\centering
\caption{Architectures and approaches evaluated in our experiments.}
\label{tab:architectures}
\setlength{\tabcolsep}{2pt}
\begin{tabular}{
>{\centering\arraybackslash}m{0.12\textwidth}
>{\centering\arraybackslash}m{0.12\textwidth}
p{0.22\textwidth}
}
\toprule
\textbf{Architecture} &
\textbf{Approach} &
\centering\arraybackslash\textbf{Operationalization in the evaluation} \\
\midrule

RAG
& \makecell[t]{OpenAI\\semantic}
& Retrieve semantically related text chunks and provide them to an LLM as grounding evidence. \\

RAG & \makecell[t]{OpenAI\\keyword}
& Retrieve text chunks using lexical matching and provide them to an LLM as grounding evidence. \\

\makecell[t]{Long\\context}
& \makecell[t]{Long-context\\prompting}
& Insert the text of all documents into the model context for every query, in the same fixed order. \\

\bottomrule
\end{tabular}
\end{table}

\vspace{-.5\baselineskip}

We used two language models of different sizes in our experiments: \texttt{gpt-5.4-mini} (version 2026-03-17) and \texttt{gpt-5.4-nano} (version 2026-03-17). We focused on smaller\footnote{Models are commonly characterized by their parameter count. The precise numbers of parameters in GPT-5.4 Mini and GPT-5.4 Nano are not publicly available.} frontier models because SME-oriented training systems must balance answer quality with low operating cost and responsive interaction. As suggested by OpenAI, ``\textit{GPT-5.4 nano is designed for tasks where speed and cost matter most like classification, data extraction, ranking, and sub-agents}'' \parencite{openai_gpt54_nano}, and ``\textit{GPT-5.4 mini brings the strengths of GPT-5.4 to a faster, more efficient model designed for high-volume workloads}" \parencite{openai_gpt54mini}.
At the time of writing, the the mini model costs \$0.75 per 1 million input tokens and \$4.50 per 1 million output tokens. 
In comparison, the nano model is significantly cheaper, priced at \$0.20 per 1 million input tokens and \$1.25 per 1 million output tokens. 
This makes the nano model's input price nearly 4 times cheaper and its output price roughly 3.6 times cheaper than the mini model. 
Both models have a context window of 400,000 tokens.

\subsection{Benchmark}
\vspace{-0.9em}

We use the benchmark introduced in \textcite{Singh2026}. 
That study developed a RAG-based multimodal manufacturing safety chatbot through a design science methodology and proposed six phases: source curation and corpus development, knowledge base and indexing, chatbot demonstration, benchmark question-answer development, automated evaluation, and human evaluation.
The benchmark was created from authoritative safety and equipment documents, including OSHA regulations and technical guidance, and original equipment manufacturer manuals for the target machines.
The authors extracted and cleaned text, segmented long documents into meaningful sections, standardized filenames for traceability, and constructed knowledge bases for multiple retrieval mechanisms.
We build on the work by \textcite{Singh2026} by evaluating AI architecture choices beyond RAG in the context of SME manufacturing training.

The benchmark includes expert-validated a number of question-answer pairs for three representative machines: a Bridgeport Manual Mill, a Haas TL-1 CNC lathe, and a Universal Robots UR5e collaborative robot. 
These machines span legacy manual equipment, automated CNC equipment, and human-robot collaborative equipment.
The original benchmark contains 51 Bridgeport questions, 51 Haas TL-1 questions, and 60 UR5e questions. 
Questions vary in difficulty, in that some can be answered from a single manual section, whereas others require integrating scattered information or interpreting safety implications. 
The original study used gold answers validated by manufacturing, occupational safety, human factors, and engineering experts.
To ensure that the input remained within the models' context windows, we restricted grounding to the manuals of the three target machines used in our experiments.

\subsection{Metrics}

The main evaluation metrics regard the two major design requirements that motivate the epistemic accuracy discussion in this paper: \textit{accuracy} and \textit{cost}. 

Accuracy is measured with two binary LLM-judge outcomes: whether the answer is \textit{correct} relative to the gold answer provided by experts and whether the answer is \textit{helpful}. 
The main variable of interest is correctness, because safety training requires operationally safe and technically accurate guidance. 
Helpfulness is reported as a complementary outcome because a partially helpful answer may still support training even when it misses a detail. 
GPT-5 Pro was used as the evaluation judge.

Cost is estimated from total token usage. 
We note that these costs should be interpreted as architecture-level estimates rather than stable procurement prices, because provider prices and accounting rules can change. 
The comparison is nevertheless meaningful because the same cost method is applied to all approaches.

Although not a primary metric in our analysis, we also measured \textit{latency}, defined as the time (in seconds) between submitting a query and receiving the complete response.
One might argue that this metric is somewhat pessimistic because, in practice, responses can be streamed to the user incrementally. 
As a result, time to first token could be a more relevant measure of perceived responsiveness. 
However, optimizing solely for time to first token can also be misleading, since the overall response may still take too long to complete despite an initially fast partial output.

We argue that the trade-off among accuracy, cost, and latency resembles the well-known blockchain trilemma, which posits that decentralized systems cannot simultaneously optimize decentralization, security, and scalability \parencite{buterin2021blockchain}. 
Instead, improvements in one dimension typically come at the expense of another. 
Analogously, in LLM-powered systems, achieving higher accuracy often requires more sophisticated models, larger context windows, or additional retrieval and reasoning steps, which increase both latency and cost when hardware resources are held constant. 
Conversely, solutions optimized for low cost and fast response times may sacrifice accuracy.

\section{Results}
\label{sec:results}

The studied AI solutions produced 972 answers: 162 questions across three approaches and two LLMs. 
We next analyze the results\footnote{All data and source code are publicly available at (the repository information has been removed from this submission to preserve anonymity during the double-blind peer-review process).} for accuracy, cost, and latency.


\subsection{Accuracy and helpfulness}

\begin{figure}[htb!]
\centering
\includegraphics[width=\columnwidth]{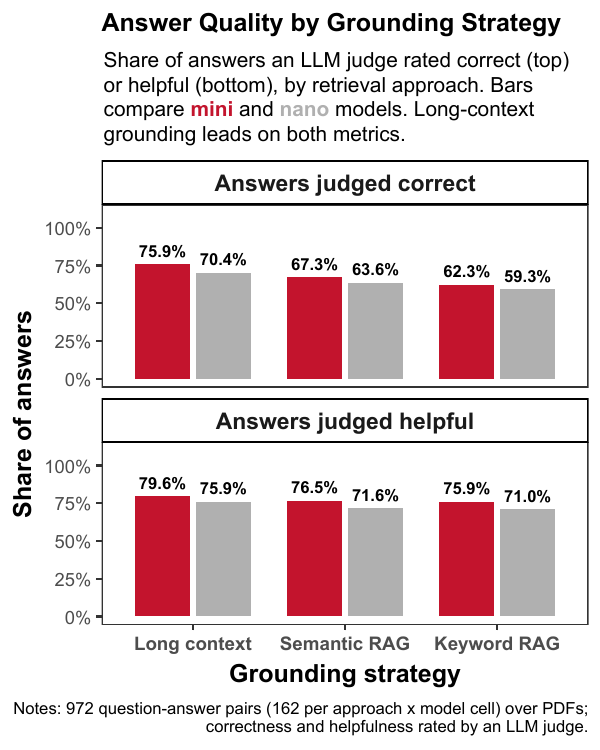}

\vspace{-0.5\baselineskip}

\caption{Answer quality by architecture \& model.}
\label{fig:accuracy_helpfulness}
\end{figure}

Figure \ref{fig:accuracy_helpfulness} summarizes the overall correctness and helpfulness results across all machines. 
Long-context prompting achieved the highest correctness for both models. 
With the mini model, long-context prompting achieved 75.93\% correctness, compared with 67.28\% for semantic RAG and 62.35\% for keyword RAG.
With the nano model, long-context prompting achieved 70.37\% correctness, compared with 63.58\% for semantic RAG and 59.26\% for keyword RAG. 
Helpfulness followed a similar pattern but with smaller separations: long-context prompting achieved 79.63\% helpful with mini and 75.93\% helpful with nano; semantic RAG achieved 76.54\% and 71.60\%; keyword RAG achieved 75.93\% and 70.99\%.

When comparing the aggregate accuracy by architecture on 324 matched question/evaluation pairs, the long-context approach achieved higher correctness at 73.1\% versus 65.4\% for OpenAI semantic RAG approach, a difference of 7.7 percentage points in favor of long context. 
The discordant-pair breakdown shows long-context prompting was correct when semantic RAG was not in 52 cases, while semantic RAG was correct when long-context prompting was not in 27 cases. 
McNemar's test indicates this difference is statistically significant (p = 0.0049 uncorrected; p = 0.0069 with Yates correction), supporting the conclusion that long-context prompting performs significantly better on this correctness measure.
This result should not be interpreted as a universal rejection of RAG; rather, it shows that retrieval loss is consequential when the entire corpus can be loaded.

The model-size comparison is more nuanced. 
Across all solutions, the model-level paired proportion test shows a statistically significant difference between  \texttt{gpt-5.4-mini} and  \texttt{gpt-5.4-nano}: mini was correct on 68.5\% of paired cases versus 64.4\% for nano, a +4.1 percentage point advantage, with McNemar p-values of 0.0124 uncorrected, 0.0175 with Yates correction.
However, when restricting the analysis to long-context prompting, mini remains directionally higher at 75.9\% versus 70.4\% for nano, but the result is no longer robustly significant: while the uncorrected McNemar p-value is borderline at 0.0495, the Yates-corrected and exact p-values are 0.0809 and 0.0784, respectively. 
Although secondary to our analysis, we note that the helpfulness results are qualitatively similar to the correctness results.

In short, model choice appears to produce a significant correctness difference overall, but within the long-context prompting approach specifically, the evidence is weaker and does not clearly support a statistically significant model difference under corrected or exact paired tests.
In other words, the result suggests that once the entire manual corpus is placed in context, the difference between these two LLMs is less important than the architectural decision to provide all evidence in context. 
As we discuss in Section \ref{sec:discussion}, this finding motivates future research on smaller and local models for SME use, especially if grounding quality can compensate for some model-capacity differences.

It is noteworthy that we found cases in which different AI systems appropriately indicated that an answer could not be determined from the documents, yet the binary LLM judge marked the response as incorrect because the gold answer assumed information outside the textual corpus. 
In a safety-critical domain, refusing to answer when evidence is insufficient is  desirable behavior. 
Ideally, evaluation should report at least two measures: gold-answer correctness and calibrated-grounding behavior. 
For example, a response that says ``I cannot determine this from the provided manual'' should receive credit when the relevant evidence is absent. 
Because these benchmark issues affect all AI approaches equally, and because benchmark revision is not the focus of this paper, we leave benchmark improvements to future work.

\subsection{Cost}

Cost is another metric in which the architectures diverge sharply, as Table \ref{tab:cost_stats} shows. 
The average cost across approaches shows that the long-context prompting approach was substantially more expensive than the RAG-based alternatives, with an overall average cost of approximately \$0.1181 across both LLM models. 
In comparison, the keyword RAG approach had an average cost of about \$0.0046, while the semantic-RAG approach averaged roughly \$0.0045. This demonstrates that RAG-based methods reduced costs by more than an order of magnitude relative to long-context inference.
When averaging across approaches, the \texttt{gpt-5.4-mini} model incurred a substantially higher overall average cost of approximately \$0.0669, whereas the \texttt{gpt-5.4-nano} model averaged about \$0.018. 

\vspace{-0.5\baselineskip}

\begin{table}[htb!]
\centering
\caption{Cost statistics by approach in USD.}
\label{tab:cost_stats}
\setlength{\tabcolsep}{2pt}
\begin{tabular}{@{}m{0.07\textwidth}cccccc@{}}
\toprule
\textbf{Method} & \textbf{gpt-5.4} & \textbf{Min} & \textbf{Mean} & \textbf{Median} & \textbf{Max} & \textbf{Glyph} \\
\midrule
Long Context & mini & 0.1860 & 0.1864 & 0.1864 & 0.1884 & \rangeglyphcompact{0.1860}{0.1864}{0.1884} \\
Long Context & nano & 0.0496 & 0.0498 & 0.0497 & 0.0506 & \rangeglyphcompact{0.0496}{0.0497}{0.0506} \\
Keyword RAG & mini & 0.0039 & 0.0072 & 0.0072 & 0.0104 & \rangeglyphcompact{0.0039}{0.0072}{0.0104} \\
Keyword RAG & nano & 0.0010 & 0.0019 & 0.0019 & 0.0032 & \rangeglyphcompact{0.0010}{0.0019}{0.0032} \\
Semantic RAG & mini & 0.0022 & 0.0070 & 0.0071 & 0.0099 & \rangeglyphcompact{0.0022}{0.0071}{0.0099} \\
Semantic RAG & nano & 0.0006 & 0.0019 & 0.0019 & 0.0029 & \rangeglyphcompact{0.0006}{0.0019}{0.0029} \\
\bottomrule
\end{tabular}

\vspace{0.25\baselineskip}

Note: The final column shows the row-normalized min--max interval; the filled point denotes the median.
\end{table}

\vspace{-0.5\baselineskip}

Overall, these results indicate that both RAG-based approaches and the nano model provide substantial cost savings compared to long-context prompting and the mini model.
These findings are expected given both the pricing structure of the models and the number of tokens processed by each approach. 
In particular, the long-context prompting approach processed approximately 247,754 input tokens on average, whereas the keyword- and semantic-based RAG approaches processed only 8,569 and 8,399 input tokens, respectively.

It is important to emphasize that the preceding analysis considers only the costs associated with processed and generated tokens. 
In other words, it does not include the additional retrieval-related costs incurred by the RAG approaches, which are more difficult to estimate precisely. 
For example, we used OpenAI's \texttt{text-embedding-3-small} model to generate embeddings for the retrieval pipeline, which costs \$0.02 per one million tokens processed. 
The resulting embeddings were then stored using OpenAI's vector storage infrastructure, priced at \$0.10 per GB per day. 
Nevertheless, we omit these retrieval-specific costs from the main analysis because they are comparatively small and do not qualitatively alter the overall conclusions reported here.

\subsection{Latency}

Table \ref{tab:elapsed_time_summary} summarizes the latency results. 
Long-context prompting had both the lowest mean and median latency of 4.69 and 2.82 seconds (sec), respectively. 
Keyword RAG had a mean of 5.09 sec and a median of 3.63 sec. Semantic RAG had a mean of 5.40 sec and a median of 4.00 sec. 
Paired t-tests were conducted to compare latency values between long-context prompting and the two RAG-based approaches. 
The comparison between long-context prompting and keyword RAG showed no statistically significant difference in latency values ($t = -0.965$, $p = 0.335$).
Similarly, the comparison between long-context prompting and semantic RAG did not reveal a significant difference ($t = -1.524$ , $p = 0.128$). 
These results suggest that the average elapsed times for the approaches were comparable across the 324 paired observations.

\vspace{-0.5\baselineskip}

\begin{table}[htb]
	\centering
	\caption{Summary statistics of latency in seconds.}
	\label{tab:elapsed_time_summary}
	\setlength{\tabcolsep}{2pt}
	\renewcommand{\arraystretch}{1.55}
	
	\begin{tabular}{p{0.09\textwidth}cccc>{\centering\arraybackslash}p{2.4cm}}
		\toprule
		\textbf{Approach} & \textbf{Min} & \textbf{Mean} & \textbf{Median} & \textbf{Max} & \textbf{Glyph} \\
		\midrule
		Long Context & 1.90 & 4.69 & 2.82 & 60.36 & \rangeglyph{1.90}{2.82}{60.36} \\[6pt]
		Keyword RAG & 2.46 & 5.09 & 3.63 & 47.48 & \rangeglyph{2.46}{3.63}{47.48} \\[6pt]
		Semantic RAG & 2.88 & 5.40 & 4.00 & 49.97 & \rangeglyph{2.88}{4.00}{49.97} \\
		\bottomrule
	\end{tabular}
\end{table}

\vspace{-0.5\baselineskip}

This result is practically important. 
A potential expectation is that long-context prompting could be slower because it processes many more tokens. 
Here, that expectation did not hold in a statistically significant way. 
The likely explanation is that end-to-end latency depends on more than token count: retrieval overhead, service behavior, caching, network variance, and answer length can all affect observed time.

\section{Discussion}
\label{sec:discussion}

This work compared RAG and long-context prompting as alternative AI architectures, with a particular focus on document-grounded manufacturing safety training systems potentially used by small and medium-sized manufacturers. 
Using an expert-validated benchmark covering three machines, two different LLMs, and three prompting/retrieval approaches, we evaluated 972 generated answers on accuracy, latency, and cost. 
The results show that long-context prompting produced the highest correctness, suggesting that loading the full manual corpus can reduce some retrieval-related failures when the relevant evidence is available in text. 
However, this accuracy gain came at a substantial cost: long-context prompting can be orders of magnitude more expensive than semantic RAG under our token-based accounting. 
Finally, latency differences were not statistically significant. 

Taken together, the accuracy and cost provide empirical support for our definition of a token tax of epistemic accuracy. 
In particular, the additional expenditure can be interpreted as the token tax paid for broader epistemic access: organizations incur higher recurring token costs in exchange for a greater likelihood that the model has access to, and can utilize, the evidence necessary to produce a correct answer. 
Our results suggest that the relationship between epistemic accuracy and cost might not be linear; modest gains in correctness may require disproportionately larger increases in token expenditure, making the architecture choice particularly consequential for resource-constrained SMEs.

On that note, our findings and associated tradeoffs have direct implications for SME-suitable AI training artifacts.
Long-context prompting produced the highest overall correctness, which matters because worker training in manufacturing is safety sensitive.
If an assistant gives an inaccurate answer about a robot safety mode, an emergency stop, or a lockout procedure, the consequence is not merely a poor user experience; it may create physical risk.
Therefore, accuracy should be the primary design requirement.
However, the cost results prevent a simple recommendation to use long-context prompting everywhere. 
A large cost increase is consequential for SMEs, especially when the assistant is envisioned as part of recurring onboarding, training, and daily support.
The broader implication is that SME-oriented AI should not be evaluated only by model accuracy.
It should be evaluated as an organizational artifact whose value depends on whether it is accurate enough to be trusted, fast enough to support training flow, and inexpensive enough to be used repeatedly.
While this trade-off is not absolute, our results suggest the existence of a practical trilemma in which optimizing all three objectives simultaneously remains challenging.
Such observations motivate many research and design directions.

First, a practical SME strategy is to adopt a hybrid architecture. For example, RAG-based systems could serve as the default because they are far less expensive, whereas long-context prompting could be reserved for cases in which RAG confidence is low, the question appears to require multi-hop synthesis, or a training module contains a small enough document set that full-context prompting is affordable.
This hybrid design aligns with the broader view in the information systems community regarding AI as an agentic artifact whose autonomy must be managed, constrained, and evaluated \parencite{Berente2021,BairdMaruping2021}.
The system should not merely answer; it should decide when to retrieve, ask for clarification, escalate to long context, and when to refuse.

Second, the comparative accuracy of the tested models is particularly important.
Within long-context prompting, the mini model outperformed the nano model, but the difference was not statistically significant.
This suggests that, for manual-grounded training, evidence availability may matter as much as model size once a baseline level of reasoning capability is present.
Future research could test whether lightweight local models, possibly fine-tuned for safety-instruction style and combined with efficient on-premise RAG, can deliver acceptable performance at lower recurring cost.
Local deployment is attractive for SMEs with privacy concerns, weak network connectivity, or limited tolerance for per-query cloud spending.
It may also support training in facilities where proprietary manuals and process details cannot leave the site.

Third, given the high costs associated with long-context prompting, future systems could explore dynamic context management.
Specifically, rather than placing all available manuals in the model context for every query, the system could first identify the relevant machine, task, or training module and then load only the corresponding manual sections.
This approach would preserve some advantages of long-context prompting while reducing unnecessary token processing.
In practice, such a strategy could combine metadata filters, lightweight classifiers, user clarification questions, and retrieval confidence scores to decide when a narrow context is sufficient and when broader context is needed.

Fourth, traditional RAG remains limited due to relying primarily on text. 
Specifically, text-only ingestion may be inadequate for manufacturing manuals because safety knowledge is often visual: connector ports, warning icons, tool attachments, robot poses, labels, and layout diagrams may not appear in extracted text. 
Emerging multimodal RAG systems offer a promising alternative \parencite{geminiembedding2native}. 
These systems can support stronger pipelines that embed videos, images, text, and audio within a shared representation space and enable the retrieval of  evidence across these modalities.

We acknowledge that our study has limitations.
In particular, the results are based on an existing benchmark rather than a live deployment with front-line manufacturing workers.
This limitation reflects the broader scope of our ongoing work, whose ultimate goal is not only to develop an AI assistant but also to embed it within a full-fledged manufacturing training environment.
In particular, both architectures produced responses within a range that could support reflective training and virtual reality (VR) guided practice.
Such a VR system can support explanation, rehearsal, and procedural learning by placing workers in a realistic but safe environment where they can practice procedures, recognize hazards, and make decisions without interrupting production or being exposed to physical risk.
However, the challenges discussed above also apply in a VR setting.
For example, the embedded assistant should retrieve not only text but also scene state, object identity, worker location, and equipment configuration.
Multimodal RAG can then connect what trainees see in VR to the relevant manual sections and safety rationale.
Our study provides an important step toward the goal of producing a manufacturing training environment that can be adopted by manufacturers of different sizes, including small and medium-sized firms.

\printbibliography

\section*{Data and Code Availability} 

Code and evaluation benchmarks are available at \url{https://github.com/fmegahed/safety_rag_evaluation}.

\section*{Acknowledgments}

This work was funded by the Ohio Bureau of Workers' Compensation Worker Safety Innovation Center
(WSIC) Grant \texttt{\#WSIC26-250206-009}. Dr. Megahed and Carvalho also received support from the \textit{Raymond
E. Glos Professorship} and the \textit{Dinesh \& Ila Paliwal Innovation Chair}. The authors also appreciate the valuable feedback provided by \textit{Engineered Profiles}, \textit{Yamaha Motor Company of North America}, and \textit{MeetKai, Inc.} on our benchmark Q\&A datasets.

\end{document}